\documentclass[preprint]{aastex}

\begin{document}

\title {Physical Nature of the Eclipsing $\delta$ Scuti Star AO Serpentis}
\author{Jang-Ho Park$^1$, Jae Woo Lee$^1$, Kyeongsoo Hong$^2$, Jae-Rim Koo$^3$, and Chun-Hwey Kim$^4$}
\affil{$^1$Korea Astronomy and Space Science Institute, Daejon 34055, Republic of Korea}
\affil{$^2$Institute for Astrophysics, Chungbuk National University, Cheongju 28644, Republic of Korea}
\affil{$^3$Department of Astronomy and Space Science, Chungnam National University, Daejeon 34134, Republic of Korea}
\affil{$^4$Department of Astronomy and Space Science, Chungbuk National University, Cheongju 28644, Republic of Korea}

\begin{abstract}
We present the absolute properties of the eclipsing binary AO Ser with a pulsating component from our $BV$ photometric and high-resolution spectroscopic observations, which were performed between April and May 2017.
The radial velocities (RVs) for both components were measured, and the effective temperature and projected rotational velocity of the primary star were determined to be $T_{\rm eff,1}$ = 8,820 $\pm$ 62 K and
$v_1 \sin i_{1}$ = 90 $\pm$ 18 km s$^{-1}$, respectively, by comparing the observed spectrum with the Kurucz models.
The accurate fundamental parameters of AO Ser were determined by a simultaneous analysis of the light and RV curves.
The masses and radii of the primary and secondary components are $M_1$ = 2.55 $\pm$ 0.09 M$_\odot$ and $R_1$ = 1.64 $\pm$ 0.02 R$_\odot$ and
$M_2$ = 0.49 $\pm$ 0.02 M$_\odot$ and $R_2$ = 1.38 $\pm$ 0.02 R$_\odot$, respectively.
Multiple frequency analyses for the eclipse-subtracted light residuals were conducted.
As a result, we detected two frequencies of $f_1$ = 21.852 days$^{-1}$ and $f_2$ = 23.484 days$^{-1}$.
The evolutionary position on the HR diagram and the pulsational characteristics indicate that the primary star is a $\delta$ Sct pulsator with a radial fundamental mode.
On the other hand, the relatively evolved secondary is oversized for its own mass.
\end{abstract}

{}

\section{Introduction}

Eclipsing binaries (EBs) provide a unique opportunity to accurately determine the fundamental parameters (e.g. mass and radius) without any assumptions from the analysis of light and radial velocity (RV) curves.
The interior structure of stars is difficult to observe directly, but it can be studied through the frequency analysis of pulsating variables.
In this respect, EBs with pulsating stars are interesting objects because they have the advantage that their binary properties can be studied together with the asteroseismic parameters.
From these studies, we can obtain very useful information for understanding the evolution and interior structure of stars.
Although a number of pulsating EBs have been found from various channels such as {\it Kepler} (Borucki et al. 2010; Koch et al. 2010) and
the Transiting Exoplanet Survey Satellite ({\it TESS}; Ricker et al. 2015),
the samples for physical properties are still insufficient, so accurate study of these objects is required.
Recently, a series of papers (Hong et al. 2015, 2017, 2019; Koo et al. 2016; Lee et al. 2018, 2020) has contributed to studying the fundamental parameters of pulsating EBs.

This paper is concerned with AO Ser ($\rm BD+17^{o} 2942$, 2MASS J15581840+1716101, TYC 1496-3-1; $V\rm_T$=$+$11.04, $(B-V)\rm_T$=$+$0.22), which was found to be an Algol-type binary star by Hoffmeister (1935).
Guthnick \& Prager (1936) reported a brightness change between 10.5 mag and 12.0 mag from photographic observations.
The spectral type of the primary star has been classified as A2$-$A7 by previous investigators (Brancewicz \& Dworak 1980; Kreiner et al. 2001; Hoffman 2009; Yang et al. 2010, hereafter YANG; Alton \& Pr\v sa 2012, hereafter AP; Hamb\'alek 2015).
The mass ratio ($q$) of this binary system was derived from photometric and spectroscopic methods.
First of all, Hoffman (2009) determined the mass ratio to be 0.178 from the double-lined RVs measured by his spectroscopic observations.
Subsequently, YANG, AP, and Hamb\'alek (2015) reported the mass ratio as 0.220, 0.235, and 0.396, respectively, by the $q$-search method based on their own photometric observations.
Chehlaeh et al. (2017) determined an accurate mass ratio of 0.183 and
the absolute parameters through a simultaneous analysis of the $BV$ light and RV curves.
Kim et al. (2004, hereafter KIM04) found an oscillation with a period and an amplitude of 1.12 h and 0.02 mag in the $B$ band, respectively, through frequency analysis.
They suggested that AO Ser is an oscillating eclipsing Algol (oEA) star (Mkrtichian et al. 2004).
Hamb\'alek (2015) detected a pulsation period  of 0.96 $\pm$ 0.05 h in his light curves by applying various analysis options.

The main purpose of this study is to accurately determine the absolute dimensions of AO Ser by simultaneously analyzing our new light curves with high-resolution spectroscopic data and
to present the pulsational characteristics through multiple frequency analyses.
The remainder of this paper is organized as follows:
The observations and data reductions are described in Section 2.
The spectral analysis for the RV measurements and the atmospheric parameters is presented in Section 3.
Section 4 describes the binary modeling and pulsation frequencies.
Lastly, the results of this study are summarized and discussed in Section 5.

\section{Observations and Data Reductions}

The CCD photometric observations of AO Ser were performed using an ARC 4K CCD camera and a $BV$ filter set attached to the 1.0 m reflector at the Mt. Lemmon Optical Astronomy Observatory (LOAO) for six nights from May 22 to 27, 2017.
The focal ratio of the Cassegrain reflector is f/7.5.
The CCD chip has 4,096$\times$4,096 pixels of 15$\mu$m size, and the field of view (FoV) is 28$\arcmin$.0$\times$28\arcmin.0.
All observed images were pre-processed using the IRAF/CCDRED package, and the aperture photometry was performed with the IRAF/PHOT package.
Considering brightness and constancy in apparent light, we selected TYC 1496-1169-1 ($V\rm_T$=$+$10.22 mag) and TYC 1496-227-1 ($V\rm_T$=$+$11.72 mag) as the comparison (C) and the check stars (K), respectively, and
the standard deviations of the (K$-$C) magnitudes was about $\pm$0.006 mag in both bands.
Finally, we obtained 2,440 and 2,434 individual data points for $B$ and $V$ bandpasses, respectively.

Spectroscopic observations were performed using the 1.8 m reflector and the Bohyunsan Optical Echelle Spectrograph (BOES) at the Bohyunsan Optical Astronomy Observatory (BOAO) for four nights between April and May 2017.
The BOES has five optical fibers covering the wavelengths range from 3,600 to 10,200 $\rm \AA$ (Kim et al. 2007).
We used the largest fiber (300 $\mu$m, $R$ = 30,000) to increase the signal-to-noise (S/N) ratio of the spectra.
The observations proceeded with an exposure time of 1500 s, and a total of 29 high-resolution spectra were obtained.
All observed spectra were pre-processed using the IRAF/CCDRED package and extracted to one-dimensional spectra with the IRAF/ECHELLE package.
The S/N ratios between 5,000 and 6,000 $\rm \AA$ were approximately 40.

\section{Spectral Analysis}

The programs FDB\textsc{inary} (Iliji\'c et al. 2004), which implements disentangling based on the Fourier-domain (Hadrava 1995), and
RaveSpan (Pilecki et al. 2013, 2015) that supports three major measurement methods (CCF, TODCOR, and BF),
were considered as a tool to obtain the RVs.
However, it was difficult to apply both programs to our observed spectra,
because the luminosity contribution of the secondary star is very low ($l_{2,V}$ $\approx$ 5 \%) and the signal was not sufficient to measure the RVs.
Since the contributions of both components to the Fe I $\lambda$4957.61 line were clearly separated in each spectrum,
the RVs of each component were measured about 5 times by two Gaussian functions using the IRAF $splot$ task.
The RVs and their errors were calculated from the average and standard deviations ($\sigma$) of the values measured by Gaussian fitting, and they are listed in Table 1.

To determine the atmospheric parameters of the primary star, we used a spectrum observed at the secondary minimum ($\phi$ = 0.507)
because the orbital inclination ($i$) of AO Ser is 90 deg and the secondary star is completely occulted by the primary star at this orbital phase (see Section 4).
First of all, we adopted four regions (Fe I $\lambda$4046, Fe I $\lambda$4260, Fe I $\lambda$4271, and Fe I $\lambda$4383) corresponding to the temperature indicators for the dwarf stars of A0$-$F0,
from the \textit{Digital Spectral Classification Atlas} of R. O. Gray\footnote{More information is available on the website: \url{https://ned.ipac.caltech.edu/level5/Gray/frames.html}}.
Then, to find the most suitable atmospheric parameters,
a total of 22,750 synthetic spectra were generated with ranges of ${T_{\rm eff} = 7,510-10,000}$ K and $v \sin i = 54-144$ km s$^{-1}$ from the ATLAS-9 atmosphere models of Kurucz (1993).
In this procedure, we assumed that the solar metallicity (Fe/H) and microturbulent velocity are 0.0 and 2.0 km s$^{-1}$, respectively, and
the surface gravity was applied to be log $g$ = 4.3 obtained from our binary parameters presented in the next section.

Finally, we found the respective values for effective temperature ($T_{\rm eff}$) and
projected rotational velocity ($v \sin i$) through a $\chi^2$ minimization of the differences between the observed spectrum and the synthetic spectra.
This is the same process that was applied by Hong et al. (2017, 2019), Lee et al. (2018, 2020), and Park et al. (2018).
The results are shown in Figure 1.
The most suitable parameters to the observed spectrum were determined to be $T_{\rm eff,1}$ = 8,820 $\pm$ 62 K and $v_1 \sin i_{1}$ = 90 $\pm$ 18 km s$^{-1}$, respectively.
Here, these errors were calculated as the standard deviations of the best-fitting parameter values obtained in each spectral region.
The four adopted spectral regions and the most suitable synthetic spectrum are presented together in Figure 2.

\section{Binary Modeling and Pulsation Frequencies}

To obtain the binary parameters of AO Ser, we analyzed our $BV$ light and RV curves acquired during the same period.
The binary modeling was performed using the 2003 version of the Wilson-Devinney synthesis code (Wilson \& Devinney 1971; van Hamme \& Wilson 2003; hereafter W-D).
The orbital period of AO Ser has varied in a combination of a downward parabola and sinusoidal variation (Khaliullina 2016).
Therefore, we collected eight primary minima between 2,457,127.5 and 2,457,898.7 including our observing interval, which have a constant orbital period.
The initial epoch ($T_0$ = HJD 2,457,127.50766 $\pm$ 0.00043) and orbital period ($P$ = 0.879349845 $\pm$ 0.000000032 day) was derived by applying a least-squares fit to these timings.
In the W-D run, the effective temperature of the primary star ($T_{\rm 1}$) was fixed at 8,820 K as determined by the spectral analysis in the previous section.
The gravity-darkening exponents ($g$) and the bolometric albedos ($A$) were adopted to be $g_{1}$ = 1.0 and $g_{2}$ = 0.32 (von Zeipel 1924; Lucy 1967), and
$A_{1}$ = 1.0 and $A_{2}$ = 0.5 (Rucinski 1969a,b) because the primary and secondary components should have radiative and convective atmospheres, respectively, according to the temperature of each component presented in Table 2.
Linear bolometric ($X$, $Y$) and monochromatic ($x$, $y$) limb-darkening coefficients were taken from van Hamme (1993).
The possible existence of a circumbinary object around AO Ser was proposed by eclipse timing analyses (YANG, AP, Khaliullina 2016).
Therefore, we searched for a possible third light ($\ell_{3}$), but its parameter converged at zero.
Thus, we set the third light to be $\ell_{3}$ = 0.0 in further analyses.

Our binary star modeling began with mode 2 (detached binary) but converged to mode 5 (semi-detached binary with the secondary filling its Roche lobe).
We adjusted the photometric (i.e., the orbital inclination $i$, the temperature of the secondary star $T_2$, the dimensionless surface potential $\Omega$, and the monochromatic luminosity of the primary star $L_1$) and
spectroscopic (i.e., the system velocity $\gamma$, the semimajor axis $a$, and the mass ratio $q$) parameters and
ran the W-D code until the corrections of each parameter became less than its standard deviation.
The final results are listed in Table 2, where $r$ (volume) is the mean volume radius calculated from the tables given by Mochnacki (1984).
Figure 3 shows the RV curves of AO Ser with model fits.
In this figure, we plotted together Hoffman's (2009) measurements for comparison, where his RVs are the result adding our system velocity ($\gamma$).
Figure 4 displays the $BV$ light curves of AO Ser with model fits.
Applying the light and velocity parameters to the JKTABSDIM code (Southworth et al. 2005), we determined the absolute dimensions of each component listed at the bottom of Table 2,
where the luminosity ($L$) and bolometric magnitudes ($M_{\rm bol}$) were calculated by adoption of the absolute values of the Sun ($T_{\rm eff}$$_\odot$ = 5,777 K and $M_{\rm bol}$$_\odot$ = +4.75) given by Zombeck (1990).
The absolute visual magnitudes ($M_{V}$) used the bolometric corrections (BCs) of Girardi et al. (2002) suitable for the effective temperatures of each component.

To investigate the pulsational characteristics of AO Ser, we used only the light curve residuals from phase 0.134 to 0.866 excluding the primary eclipse.
According to our binary model, the pulsating light curves during the primary eclipse may be distorted because the secondary component partially blocked the pulsating primary star.
Multiple frequency analyses were performed using the discrete Fourier transform program PERIOD04 (Lenz \& Breger 2005).
Through an iterative pre-whitening procedure,
we detected two frequencies of $f_1$ = 21.852 days$^{-1}$ and $f_2$ = 23.484 days$^{-1}$ with an S/N amplitude ratio higher than the empirical criterion of 4.0 proposed by Breger et al. (1993) in the $B$ band.
The results and their uncertainties calculated by the equations in the paper of Montgomery \& O'Donoghue (1999) are listed in Table 3.
The spectral window and the amplitude spectra of the $B$ band residuals are presented in Figure 5.
Figure 6 displays the phase-folded light variations with the synthetic curves calculated from each frequency.

\section{Summary and Discussion}

In this paper, $BV$ photometric and high-resolution spectroscopic observations of the pulsating EB AO Ser were presented.
We measured the RVs at the Fe I $\lambda$4957.61 region where the two components were isolated in the observed spectra.
From the semi-amplitudes ($K_1$, $K_2$), the spectroscopic mass ratio was calculated to be $q = 0.191 \pm 0.004$.
The atmosphere parameters of the pulsating primary star were determined to be $T_{\rm eff,1}$ = 8,820 $\pm$ 62 K and
$v_1 \sin i_{1}$ = 90 $\pm$ 18 km s$^{-1}$, respectively.
From the simultaneous analysis of the light and RV curves, the binary parameters of AO Ser were obtained with $i$ = 90 deg, temperature difference ($\Delta T$) = 3,988 K, and $a$ = 5.59 R$_\odot$, respectively.
The absolute parameters of each component were determined as follows:
$M_1$ = 2.55 $\pm$ 0.09 M$_\odot$, $R_1$ = 1.64 $\pm$ 0.02 R$_\odot$ and $M_2$ = 0.49 $\pm$ 0.02 M$_\odot$, $R_2$ = 1.38 $\pm$ 0.02 R$_\odot$.

In Figure 7, the evolutionary positions of each component on the HR diagram are marked by star symbols, together with those of 38 other semi-detached EBs with a $\delta$ Sct component (Kahraman Ali\c cavu\c s et al. 2017).
Here, it can be seen that the primary component belongs to the main-sequence star and is near the blue edge of the $\delta$ Sct instability region.
On the other hand, the companion is outside the main-sequence stage, and is located in the region where the secondaries of semi-detached EBs with $\delta$ Sct components are distributed.
The Roche geometry of AO Ser indicated that the secondary component fills its inner lobe completely, while the hotter primary fills only 61\%.
Considering the physical properties of AO Ser,
it can be inferred that the primary was formed by mass flowed from the more massive component that was relatively evolved in the past,
while the donor lost most of its own mass and became the oversized secondary with a low mass (Sarna 1993; Erdem \& \"Ozt\"urk 2014).

From the pulsation frequency analysis, we detected two frequencies $f_1$ = 21.852 days$^{-1}$ and $f_2$ = 23.484 days$^{-1}$, respectively,
where $f_1$ is very close to $f$ = 21.5 days$^{-1}$ found by KIM04.
We calculated the pulsation constant ($Q$) of this dominant frequency $f_1$ by applying our accurate fundamental parameters to the following equation:
$\log Q = -\log f + 0.5 \log g + 0.1M_{\rm bol} + \log T_{\rm eff} - 6.456$ (Breger 2000).
The pulsation constant was calculated to be $Q$ ($f_1$) = 0.0349 $\pm$ 0.0028 days,
which can be identified as the radial fundamental mode of the $\delta$ Sct stars from comparing with the theoretical model given by Fitch (1981).
On the other hand, the newly found $f_2$ has low amplitude and can suffer from 1 day$^{-1}$ aliasing.
If $f_2$ is a real pulsation frequency of AO Ser, the high frequency ratio $f_1$/$f_2$ = 0.931 indicates that $f_2$ might be a nonradial $p$-mode.

\acknowledgements We are grateful to the anonymous referee for valuable comments, which helped us to improve the manuscript.
The authors wish to thank S.-L. Kim for helpful comments on pulsation analysis.
This work was supported by KASI (Korea Astronomy and Space Science Institute) grant 2020-1-834-00.
The work by K. Hong, J.-R. Koo, and C.-H. Kim was supported by the grant Nos. 2017R1A4A1015178 and 2019R1I1A1A01056776, 2017R1A6A3A01002871, and 2020R1A4A2002885 of the National Research Foundation (NRF) of Korea, respectively.
We have used the Simbad database maintained at CDS, Strasbourg, France.

\clearpage
\begin{figure}
\centering
\includegraphics[width=1\columnwidth]{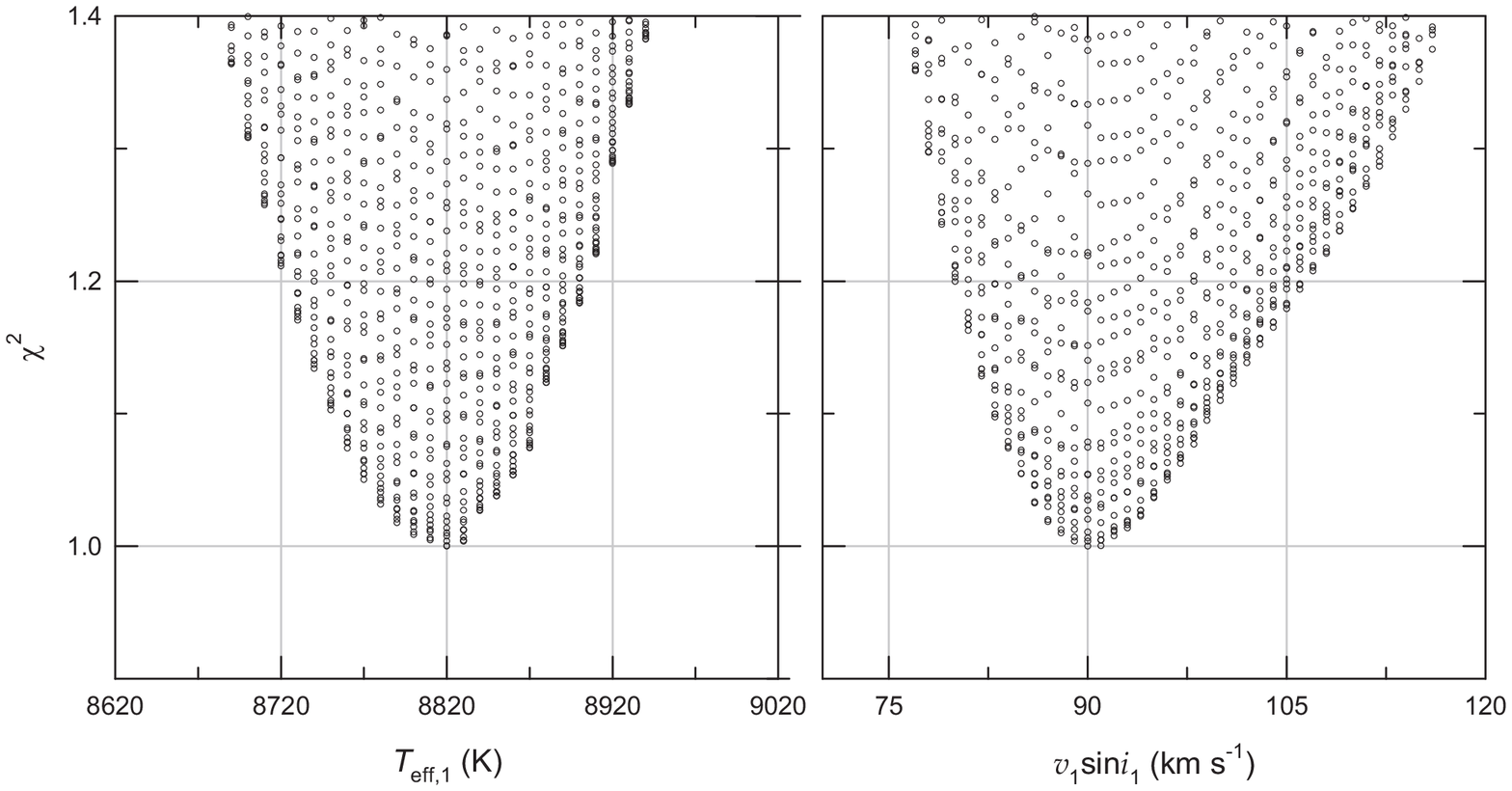}
\caption{$\chi^2$ diagrams of the effective temperature ($T_{\rm eff,1}$) and the projected rotational velocity ($v_1 \sin i_{1}$) of the primary star.}
\label{Fig1}
\end{figure}

\begin{figure}
\centering
\includegraphics[]{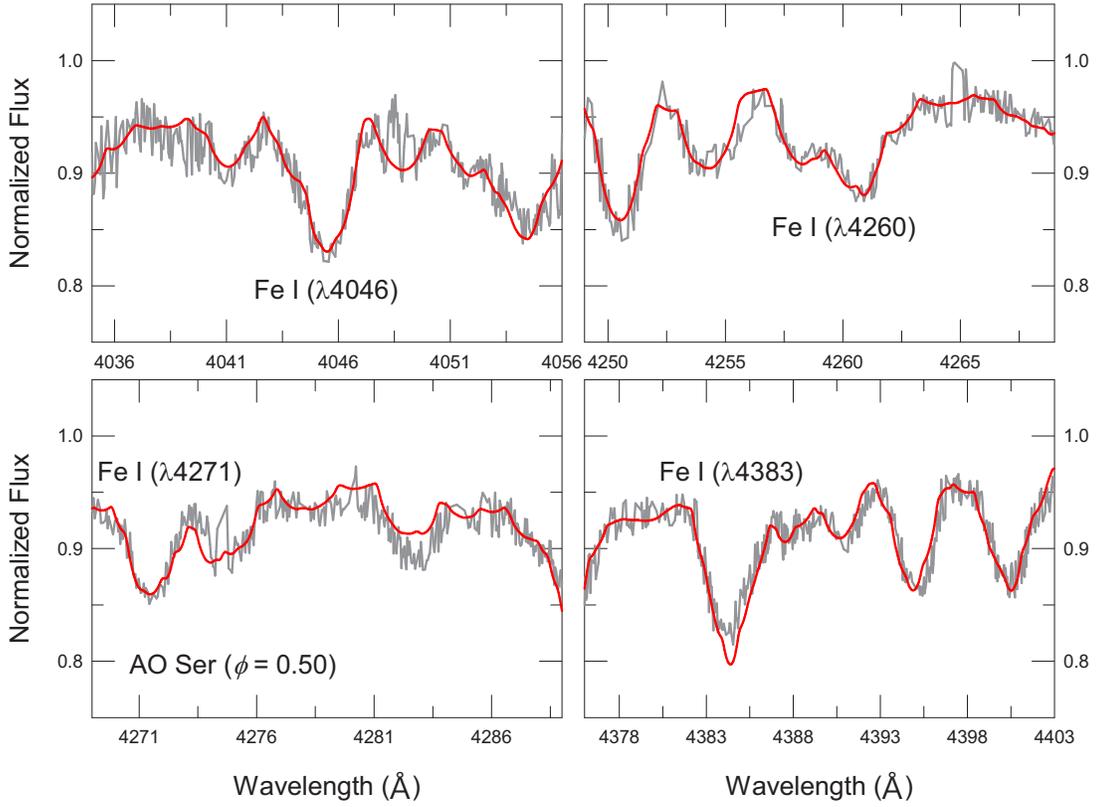}
\caption{Four spectral regions of the primary star, observed at an orbital phase of $\phi$ = 0.507 (HJD 2,457,847.2616).
The red line represents the synthetic spectrum with the best-fit parameters of $T_{\rm eff,1}$ = 8,820 K and $v_1 \sin i_{1}$ = 90 km s$^{-1}$.}
\label{Fig2}
\end{figure}

\begin{figure}
\centering
\includegraphics[]{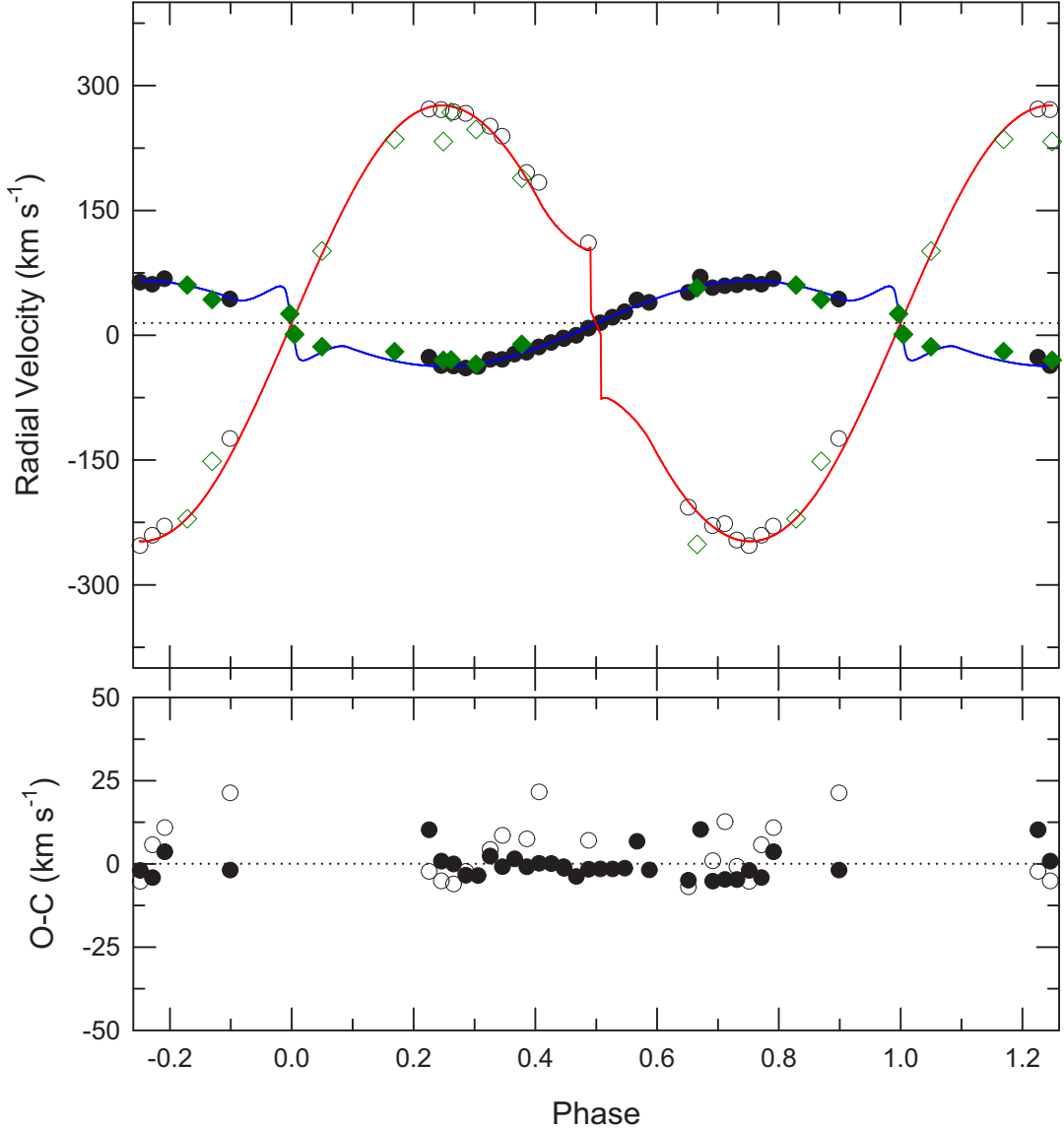}
\caption{RV curves of AO Ser with fitted models. The filled and open circles represent our double-lined RV measurements for
the primary and secondary components, respectively. The diamonds represent the double-lined RVs of
Hoffman (2009), respectively. In the upper panel, the solid curves represent the results
from a consistent light and RV curve analysis with the W-D code. The dotted line represents the system velocity of
$+$14.2 km s$^{-1}$. The lower panel shows the differences between observations and theoretical models.}
\label{Fig3}
\end{figure}

\begin{figure}
\centering
\includegraphics[]{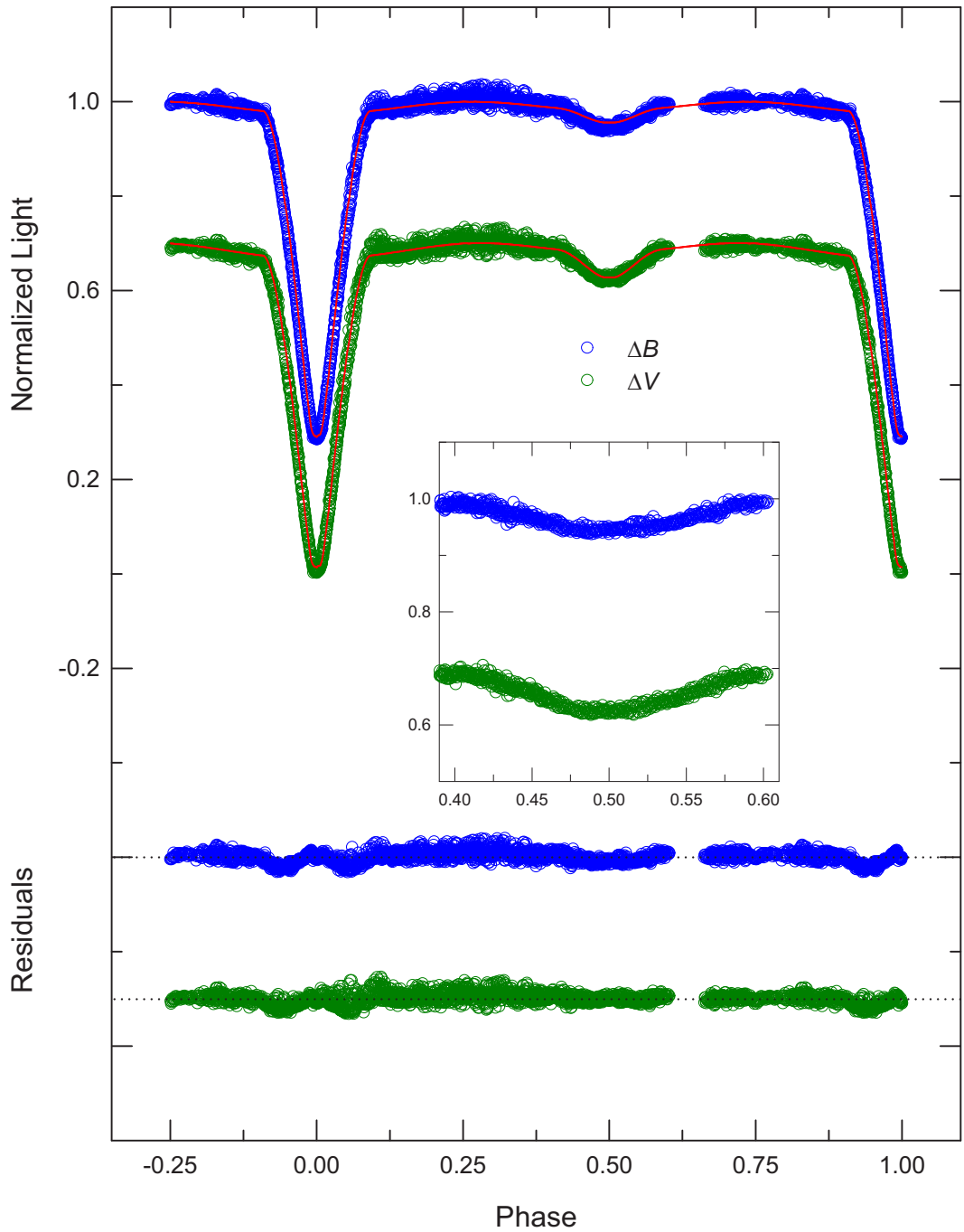}
\caption{$BV$ light curves of AO Ser with fitted models.
The circles are individual measurements taken from our LOAO observations, and the solid lines represent the synthetic curves obtained with the W-D run.
The light residuals for $B$ and $V$ are shown in the bottom, respectively.
The inset box shows the light curve from phase 0.39 to 0.61, where it can be seen that the secondary minimum is flat bottomed.}
\label{Fig4}
\end{figure}

\begin{figure}
\centering
\includegraphics[]{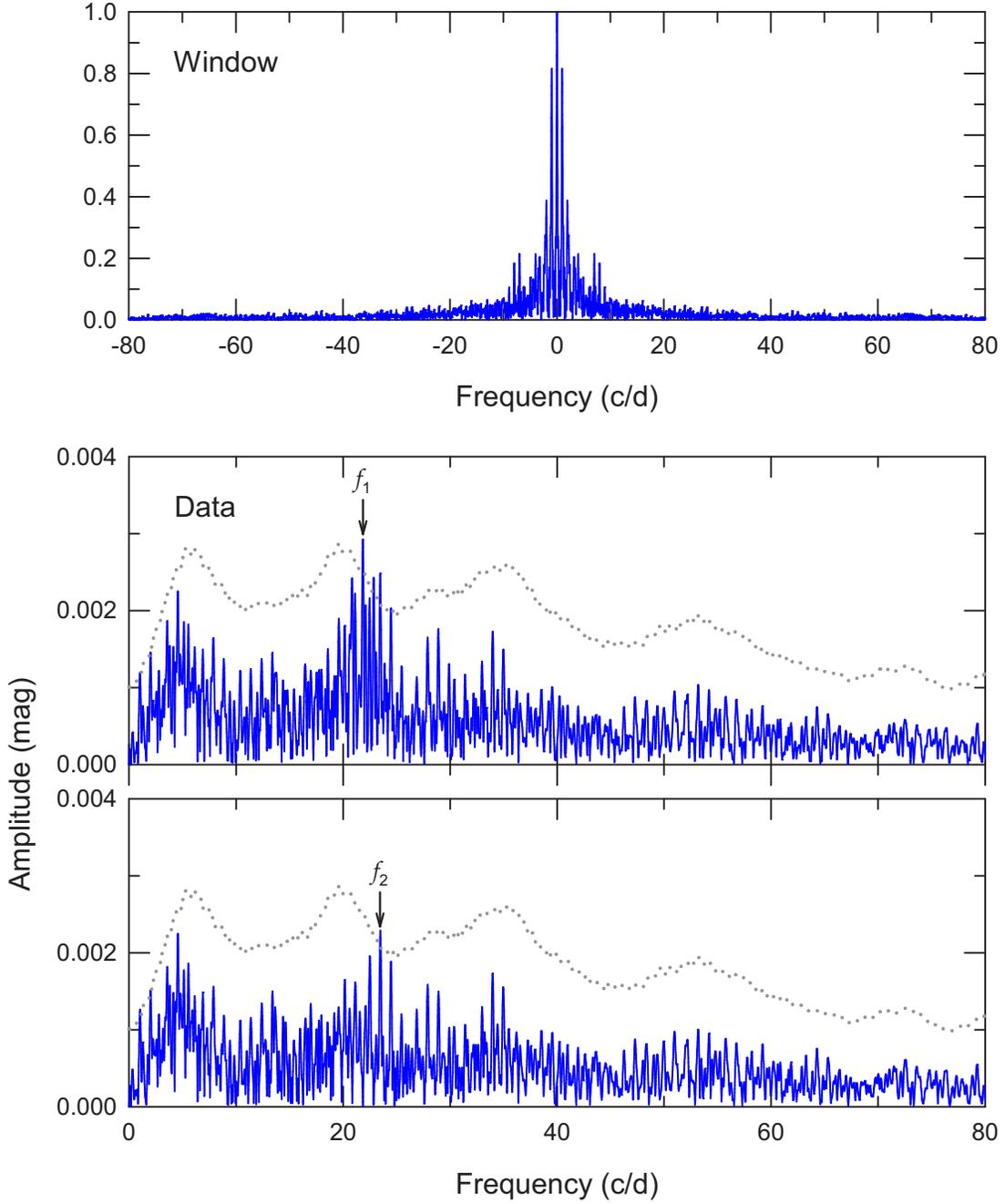}
\caption{Power spectra of the $B$-band residuals in the out-of-primary eclipsing phases.
The window spectrum is displayed in the top panel. We could detect two frequencies ($f_1$, $f_2$) in the next two panels.
Here, the dotted line represents the significance limit (S/N $\geq$ 4.0) proposed by Breger et al. (1993).}
\label{Fig5}
\end{figure}

\begin{figure}
\centering
\includegraphics[]{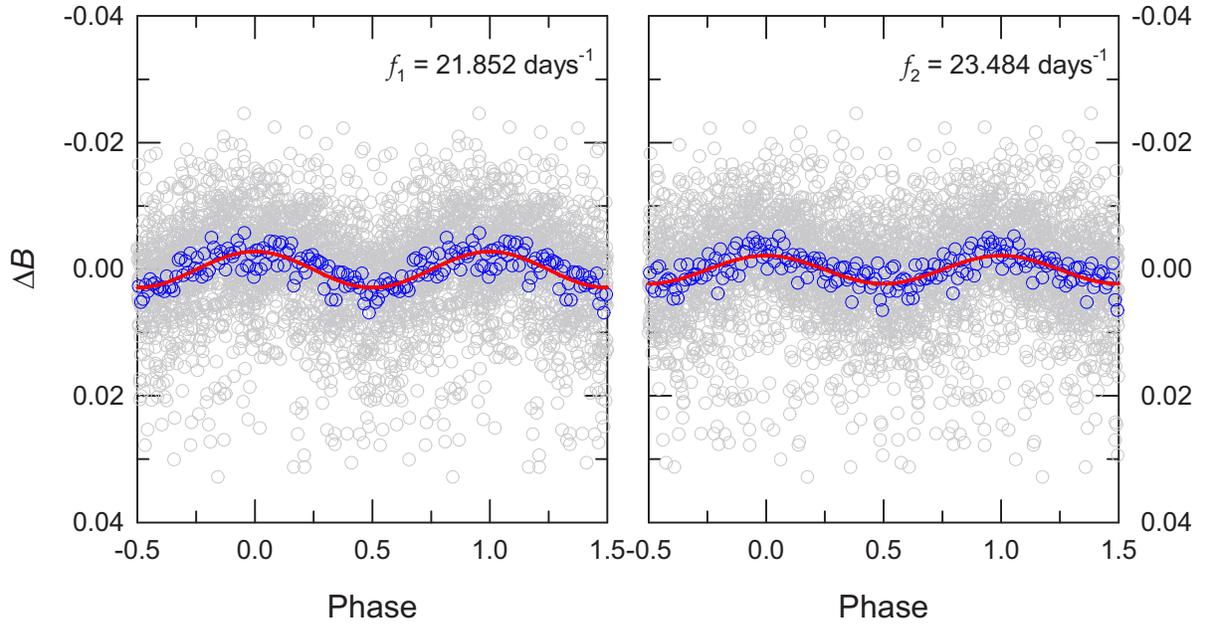}
\caption{The phase-folded light variations of the $B$-band residuals.
Blue open circles denote the representative points of residuals binned at phase 0.01 intervals, and the synthetic curves were calculated from multiple frequency analyses.}
\label{Fig6}
\end{figure}

\begin{figure}
\centering
\includegraphics[]{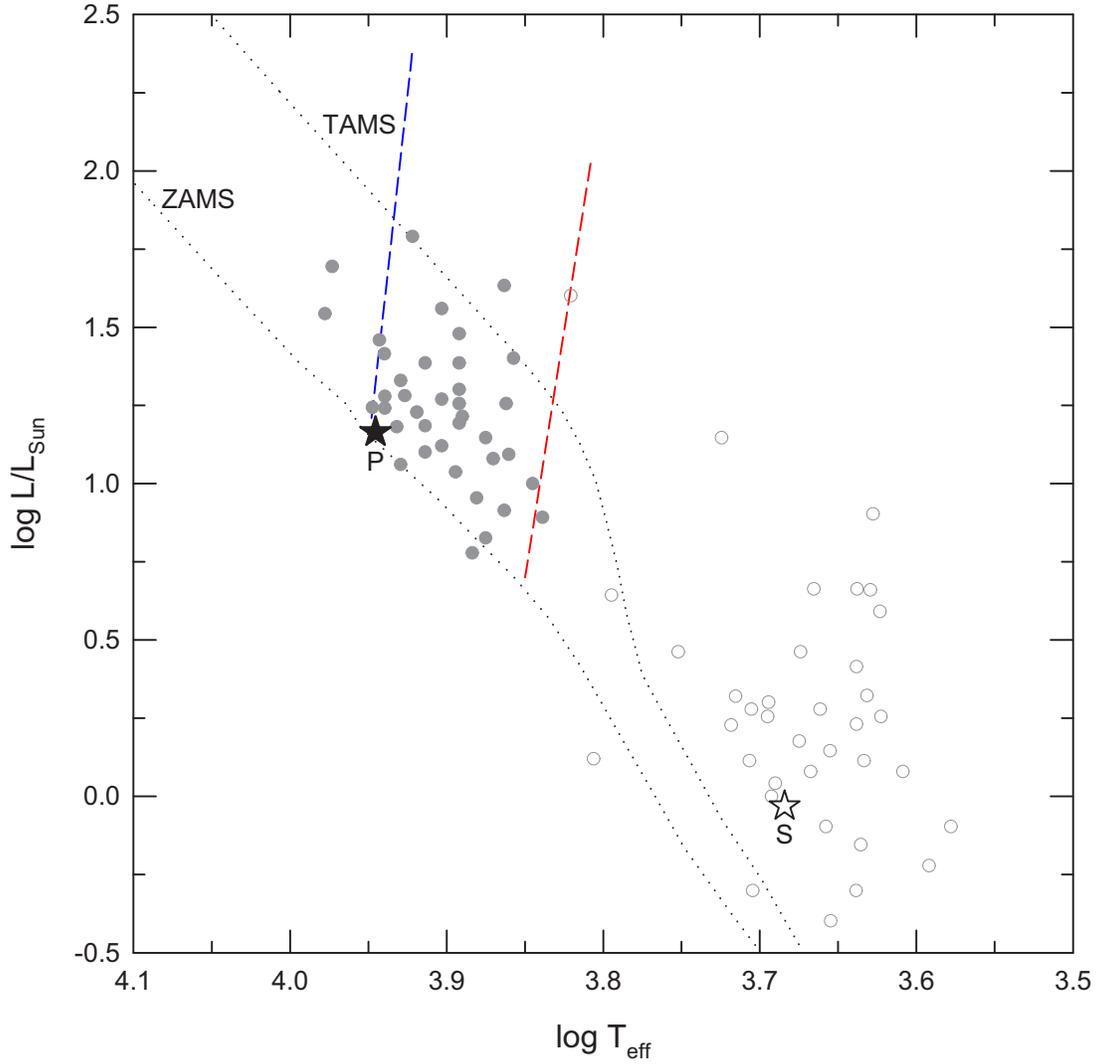}
\caption{Positions on the HR diagram for AO Ser (star symbols) and other semi-detached EBs with $\delta$ Sct component (circles, Kahraman Ali\c cavu\c s et al. 2017).
The filled and open circles represent the primary and secondary stars, respectively.
The dashed blue and red lines represent the theoretical edges of the $\delta$ Sct instability strip (Rolland et al. 2002; Soydugan et al. 2006).}
\label{Fig7}
\end{figure}

\clearpage
\begin{deluxetable}{lcrrrr}
\tabletypesize{\scriptsize}
\tablewidth{0pt}
\tablecaption{Radial Velocities of AO Ser.}
\tablehead{
\colhead{HJD}          & \colhead{Phase}        & \colhead{$V_{1}$}      & \colhead{$\sigma_{1}$} & \colhead{$V_{2}$}      & \colhead{$\sigma_{2}$} \\
\colhead{(2,450,000+)} &                        & \colhead{km s$^{-1}$}  & \colhead{km s$^{-1}$}  & \colhead{km s$^{-1}$}  & \colhead{km s$^{-1}$}
}
\startdata
7,846.1346             &  0.226                 &  $-$26.6               &      1.7               &    271.7               &      2.3               \\
7,846.1523             &  0.246                 &  $-$36.5               &      1.8               &    271.0               &     16.5               \\
7,846.1699             &  0.266                 &  $-$37.1               &      1.1               &    268.4               &      4.0               \\
7,846.1875             &  0.286                 &  $-$39.5               &      1.3               &    266.6               &      4.3               \\
7,846.2052             &  0.306                 &  $-$37.6               &      2.1               &  \dots                 &  \dots                 \\
7,846.2228             &  0.326                 &  $-$29.2               &      1.4               &    251.1               &      8.2               \\
7,846.2404             &  0.346                 &  $-$29.0               &      0.7               &    239.0               &     11.4               \\
7,846.2581             &  0.366                 &  $-$22.7               &      2.1               &  \dots                 &  \dots                 \\
7,846.2757             &  0.386                 &  $-$20.5               &      0.8               &    195.5               &      1.1               \\
7,846.2933             &  0.406                 &  $-$14.3               &      1.0               &    183.7               &      7.8               \\
7,846.3110             &  0.426                 &   $-$8.8               &      0.9               &  \dots                 &  \dots                 \\
7,846.3286             &  0.446                 &   $-$3.7               &      0.9               &  \dots                 &  \dots                 \\
7,847.2087             &  0.447                 &   $-$4.0               &      1.1               &  \dots                 &  \dots                 \\
7,847.2264             &  0.467                 &   $-$0.2               &      0.6               &  \dots                 &  \dots                 \\
7,847.2440             &  0.487                 &      8.4               &      0.7               &    110.9               &      1.4               \\
7,847.2616             &  0.507                 &     15.0               &      0.9               &  \dots                 &  \dots                 \\
7,847.2792             &  0.527                 &     21.5               &      0.7               &  \dots                 &  \dots                 \\
7,847.2969             &  0.547                 &     28.0               &      1.9               &  \dots                 &  \dots                 \\
7,847.3145             &  0.567                 &     42.1               &      1.9               &  \dots                 &  \dots                 \\
7,847.3321             &  0.588                 &     39.3               &      0.8               &  \dots                 &  \dots                 \\
7,900.1495             &  0.652                 &     51.2               &      0.9               & $-$206.8               &      8.9               \\
7,900.1670             &  0.672                 &     69.9               &      5.0               &  \dots                 &  \dots                 \\
7,900.1846             &  0.692                 &     57.0               &      1.4               & $-$228.9               &      5.3               \\
7,900.2022             &  0.712                 &     59.5               &      1.4               & $-$226.8               &      5.9               \\
7,900.2197             &  0.731                 &     60.6               &      0.4               & $-$246.3               &      2.4               \\
7,900.2373             &  0.751                 &     63.7               &      1.3               & $-$253.0               &      3.8               \\
7,900.2548             &  0.771                 &     61.2               &      1.6               & $-$240.5               &      2.9               \\
7,900.2724             &  0.791                 &     67.7               &      9.7               & $-$229.7               &      1.2               \\
7,901.2462             &  0.899                 &     43.3               &      0.9               & $-$124.1               &      0.9               \\
\enddata
\end{deluxetable}

\clearpage
\begin{deluxetable}{lcc}
\tablewidth{0pt}
\tablecaption{Binary Parameters of AO Ser.}
\tablehead{
\colhead{Parameter}                      & \colhead{Primary}   & \colhead{Secondary}
}
\startdata
$T_0$ (HJD)                              & \multicolumn{2}{c}{2,457,127.5076 $\pm$ 0.0041}   \\
$P$ (day)                                & \multicolumn{2}{c}{0.8793496 $\pm$ 0.0000047}     \\
$i$ (deg)                                & \multicolumn{2}{c}{90.0 $\pm$ 1.5}                \\
$T$ (K)                                  & 8,820 $\pm$ 62      & 4,832 $\pm$ 34              \\
$\Omega$                                 & 3.645 $\pm$ 0.002   & 2.210                       \\
$\Omega_{\rm in}$                        & \multicolumn{2}{c}{2.210}                         \\
$l_1$/($l_{1}$+$l_{2}$){$_{B}$}          & 0.9787 $\pm$ 0.0004 & 0.0213                      \\
$l_1$/($l_{1}$+$l_{2}$){$_{V}$}          & 0.9527 $\pm$ 0.0007 & 0.0473                      \\
$r$ (pole)                               & 0.2889 $\pm$ 0.0002 & 0.2298 $\pm$ 0.0002         \\
$r$ (point)                              & 0.2960 $\pm$ 0.0002 & 0.3373 $\pm$ 0.0009         \\
$r$ (side)                               & 0.2932 $\pm$ 0.0002 & 0.2390 $\pm$ 0.0002         \\
$r$ (back)                               & 0.2950 $\pm$ 0.0002 & 0.2713 $\pm$ 0.0002         \\
$r$ (volume)                             & 0.2924 $\pm$ 0.0002 & 0.2461 $\pm$ 0.0002         \\
\multicolumn{3}{l}{Spectroscopic:}                                                           \\
$a$ (R$_\odot$)                          & \multicolumn{2}{c}{5.59 $\pm$ 0.05}               \\
$\gamma$ (km s$^{-1}$)                   & \multicolumn{2}{c}{14.22 $\pm$ 0.71}              \\
$K_{1}$ (km s$^{-1}$)                    & \multicolumn{2}{c}{51.6 $\pm$ 1.1}                \\
$K_{2}$ (km s$^{-1}$)                    & \multicolumn{2}{c}{270.3 $\pm$ 3.6}               \\
$q$                                      & \multicolumn{2}{c}{0.191 $\pm$ 0.004}             \\
rms (km s$^{-1}$)                        & 4.0                 & 9.6                         \\ \hline
\multicolumn{3}{l}{Absolute Parameters:}                                                     \\
$M$ (M$_\odot$)                          & 2.55    $\pm$ 0.09  & 0.49    $\pm$ 0.02          \\
$R$ (R$_\odot$)                          & 1.64    $\pm$ 0.02  & 1.38    $\pm$ 0.02          \\
$\log$ $L/L_\odot$                       & 1.16    $\pm$ 0.02  & $-$0.03 $\pm$ 0.02          \\
$\log$ $g$ (cgs)                         & 4.42    $\pm$ 0.01  & 3.85    $\pm$ 0.01          \\
$M_{\rm bol}$ (mag)                      & $+$1.84 $\pm$ 0.04  & $+$4.83 $\pm$ 0.04          \\
BC (mag)                                 & $-$0.04             & $-$0.34                     \\
$M_{\rm V}$ (mag)                        & $+$1.88 $\pm$ 0.03  & $+$5.17 $\pm$ 0.05          \\
\enddata
\end{deluxetable}

\clearpage
\begin{deluxetable}{ccccc}
\tabletypesize{\scriptsize}
\tablewidth{0pt}
\tablecaption{Results of Multiple Frequency Analyses of the $B$-band Residuals.}
\tablehead{
                       & \colhead{Frequency}     & \colhead{Amplitude}    & \colhead{Phase}        & \colhead{S/N}          \\
                       & \colhead{(days$^{-1}$)} & \colhead{(mag)}        & \colhead{(rad)}        &
}
\startdata
$f_{1}$                & 21.852   $\pm$ 0.008    & 0.00286 $\pm$ 0.00023  & 1.71 $\pm$ 0.08        & 4.48                   \\
$f_{2}$                & 23.484   $\pm$ 0.011    & 0.00224 $\pm$ 0.00023  & 4.06 $\pm$ 0.11        & 4.30                   \\
\enddata
\end{deluxetable}

\end{document}